\documentclass[runningheads]{cls/llncs}

\usepackage[T1]{fontenc}

\usepackage{graphicx}
\graphicspath{ {./images/} }
\usepackage{enumitem}
\usepackage{xcolor}
\usepackage{soul}
\usepackage{hyperref}
\usepackage{float}

\begin{document}

    \title{NordIQuEst: the Nordic--Estonian Quantum Computing e--Infrastructure Quest}
    
    %
    %
    \author{Costantino Carugno\inst{1}\inst{4}\and
Jake Muff\inst{1}\inst{2}\and\\
Mikael P. Johansson\inst{2}\and
Sven Karlsson\inst{3}\and
Alberto Lanzanova\inst{2}
}

\institute{VTT Technical Research Center (Finland)\and
CSC -- IT Center for Science (Finland)\and
Technical University of Denmark (Denmark)\and
DEIB, Politecnico di Milano (Italy)
}

\authorrunning{C. Carugno et al.}
\titlerunning{NordIQuEst}
    
    \maketitle     

    %
    %
    \begin{abstract}
This paper presents the Nordic--Estonian Quantum Computing e--Infrastructure Quest -- NordIQuEst -- an international collaboration of scientific and academic organizations from Denmark, Estonia, Finland, Norway, and Sweden, working together to develop a hybrid High--Performance and Quantum Computing (HPC+QC) infrastructure. The project leverages existing and upcoming classical high--performance computing and quantum computing systems, facilitating the development of interconnected systems. Our effort pioneers a forward--looking architecture for both hardware and software capabilities, representing an early--stage development in hybrid computing infrastructure. Here, we detail the outline of the initiative, summarizing the progress since the project outset, and describing the framework established. Moreover, we identify the crucial challenges encountered, and potential strategies employed to address them.


\end{abstract}
    
    %
    %
    \setcounter{footnote}{0}
    \section{Introduction}
Over the years, HPC systems have served as the foundation for conducting demanding calculations and simulations. Increasingly powerful supercomputers have been deployed with elaborated software stacks and dedicated resource management~\cite{Navaux_Lorenzon_Serpa_2023}. A supercomputer is realized through a computer cluster, equipped with a specialized set of software and interconnected hardware that enables massively parallel and distributed computing, leveraging the power of many individual computing units. Recently, QC has emerged as an alternative computational paradigm, advocating the use of quantum devices in order to carry on specialized computation in a radically different way, by leveraging quantum effects~\cite{DiVincenzo_2000}. The availability of first-generation quantum computers, named ``Noisy Intermediate-Scale Quantum'' (NISQ) devices~\cite{Preskill_2018}, has allowed researchers to experiment with a variety of quantum algorithms under practical, although limited, conditions. Presently, these devices, characterized by their ability to handle from a few to a few thousand ``noisy'' qubits, face the issue of preserving information encoded in a quantum system for the required time necessary to carry out computationally difficult tasks~\cite{Smith:2022htg}. However, current improvements in technology and future efforts towards developing ``Fault--Tolerant Quantum Computers'' (FTQC)~\cite{shor1997faulttolerant}, hold the potential to overcome this issue, and solve problems that are currently intractable for classical supercomputers alone~\cite{doi:10.1126/science.1231930}. From this, efforts towards integration of HPC systems with QC have sprung forth, with the aim of ultimately addressing uniquely complex scientific and engineering problems. 

The NordIQuEst project\footnote{\href{https://nordiquest.net/}{https://nordiquest.net/}} represents a pioneering initiative to leverage the complementary strengths of both HPC and QC computing systems, blending these two different forms of computing into a hybrid HPC+QC infrastructure\cite{NordIQuEst-ERCIM}. This collaborative effort, promoted by the Nordic e--Infrastructure Collaboration NeIC\footnote{\href{https://neic.no/}{https://neic.no/}}, and brought forward by various organizations within the Nordic nations -- DTU\footnote{\href{https://www.dtu.dk/english}{https://www.dtu.dk/english}} (Denmark), VTT\footnote{\href{https://www.vttresearch.com/en}{https://www.vttresearch.com/en}} and CSC\footnote{\href{https://www.csc.fi/}{https://www.csc.fi/}} (Finland), Simula\footnote{\href{https://www.simula.no/}{https://www.simula.no/}} and Sintef\footnote{\href{https://www.sintef.no/en/}{https://www.sintef.no/en/}} (Norway), Chalmers--WACQT\footnote{\href{https://www.chalmers.se/en/centres/wacqt/}{https://www.chalmers.se/en/centres/wacqt/}} (Sweden), and University of Tartu\footnote{\href{https://ut.ee/en/home}{https://ut.ee/en/home}} (Estonia) -- aims to set up a prototypical infrastructure, incorporating HPC and QC devices, and specialized software. Integrating quantum computing into the existing framework of HPC systems introduces several challenges, including the need for compatible software environments, efficient management of resources, and robust access control \cite{mikael_p_johansson_2021_5547408}. NordIQuEst tackles these issues by developing a cohesive ecosystem that is approachable and reliable, in order to enable researchers to execute smoothly on both real QC devices and the hybrid system. In addition to preparing the infrastructure, NordIQuEst has organized several events and workshops aimed at disseminating knowledge and educating experimenters on HPC and QC topics\footnote{\href{https://enccs.github.io/nordiquest-workshop/}{NordIQuEst workshop 2022}}\footnote{\href{https://nordiquest.net/_posts/2022-12-13-Nordiquest_QC_Norway/}{Quantum Computing Norway 2022}}\footnote{\href{https://nordiquest.net/_posts/2023-12-15-QAS2023/}{Quantum Autumn School 2023}}\footnote{\href{https://nordiquest.net/_posts/2023-01-30-Young_Ambassadors/}{Norway Young Ambassadors Program}}. The project reflects the Nordic tradition of collaboration in innovation and scientific inquiry, and seeks to establish a blueprint for the integration of quantum and classical computing resources.

This paper outlines the foundational principles of HPC and QC, examines the potential for their combination, and describes the implementation of the NordIQuEst platform. It addresses the technical, operational, and conceptual challenges of merging these computing paradigms, and shares the project's approach and vision for navigating current and upcoming obstacles. This work lays the foundation for further advancement of integrated HPC+QC computing technologies, upon which future projects may leverage and expand upon.
    \section{HPC and QC}
HPC systems are advanced computing systems that leverage the aggregation and orchestration of computational resources, in order to tackle problems that demand vast amounts of computing power. Such infrastructures are constituted by the combined integration of several technologies, of which a supercomputer, in the form of a computer cluster, stands as the foundational element. A cluster is a physical infrastructure that groups many computing nodes -- individual computing resources -- that serve as a unified computational resource. The nodes are connected through a high--speed network and specialized interconnect hardware, specifically engineered to meet computational demands, leveraging the combined capabilities of one or more CPUs, GPUs, or other dedicated hardware. In addition, supercomputers are equipped with the memory and the storage necessary to accommodate the data. The supercomputer is managed on--site and is accessed remotely by users, that carry out intensive calculations using distributed parallel computing, a computational paradigm that consists in dividing the problem into smaller tasks, distributing and executing them concurrently across the nodes. Moreover, a group of tasks can be collected and processed simultaneously in batches, further reducing processing time. The efficiency of HPC systems heavily relies on the software and algorithms used, as optimizing execution for parallel processing is critical for maximizing performance. Workload management and job scheduling software is utilized, to allocate the computational resources needed without hindering overall system performance. Furthermore, a dedicated system to manage user identity and permission levels is essential, in order to properly assign computational quotas, and impose security restrictions, averting misuse of the system. Lastly, handling large volumes of data necessitates an advanced storage systems, such as a parallel file system, designed for big data storage, fast data retrieval, and high reliability.

QC represents a paradigm shift in our approach to computational problems, introducing principles that fundamentally diverge from classical computing.\cite{wendin2023quantum} At its core, quantum computing encodes information in a quantum system, and leverages phenomena from quantum mechanics -- superposition, interference, and entanglement -- to carry out the calculation. The fundamental unit of quantum information is the qubit, analogous to the bit in classical computing. Unlike a classical bit, which can take either one of two values -- $0$ or $1$ -- a qubit can exist in a superposition of both states, allowing, in principle, a quantum computer with $n$ qubits to represent and process $2^n$ possible qubit states simultaneously. In classical computers, the information is represented as a string of $n$ bits -- a bitstring -- which is inserted in a processor register, and after certain logic operations are performed, the resulting bitstring is obtained. In quantum computing, information is encoded in $n$ qubits -- a quantum state -- symbolized as a quantum register but realized in a physical system. The qubits are then manipulated using quantum gates, i.e., basic logic operations, until they reach a final state composed of a superposition of the qubits states. By measuring the qubits, the final state collapses to a classical state, a bitstring. However, as the final state is the result of a superposition of the qubit states, a single measurement yields only one possible bitstring, obtained according to its specific probability, dictated by quantum mechanics. In order to reconstruct the probability distribution of all the outcoming bitstrings in the final state, the measurement process needs to be repeated several times, which is often referred to as ``shots''. In fact, quantum interference, the ability of quantum states to amplify or cancel each other, plays a critical role during quantum computation, guiding the state towards the correct solutions by enhancing probabilities associated with desired outcomes, while suppressing the others. Similarly to how a series of logical gates designed to perform specific operations are assembled into circuits, a series of quantum gates are assembled in quantum circuits, which are unitary transformations that enable superposition and entanglement, a phenomenon where the state of one qubit is correlated with the state of another, such that the measurement of one immediately defines the state of the other. 

A given problem is thus solved by QC in a radically different way, with a quantum algorithm that pre--processes the data, encodes it in a qubit register, executes a quantum circuits several times, and post--processes the outcome. A diverse set of quantum algorithms has been conceived over the years, aimed at addressing issues that are difficult to solve using classical methods. It is important to note that quantum computers and algorithms will not replace classical computers and algorithms. In fact, one should think of QC devices as more closely resembling hardware accelerators, special--purpose solvers, rather than entire classical computers replacements. The concept of having dedicated hardware to perform specific demanding computational operations is not new to computer science. Initially conceived to advance computer graphics and image processing, Graphics Processing Units (GPUs) have proven useful in a diverse set of applications, from training neural networks to mining cryptocurrency, due to their capacity of handling intensive matrix calculations, and they are now a common component of HPC systems. Furthermore, while Application--Specific Integrated Circuits (ASICs) are custom--designed circuits, used in HPC for tasks that require high efficiency and performance, Field--Programmable Gate Arrays (FPGAs) have a versatile internal logic, and are utilized in HPC for their flexibility, and their ability to be reprogrammed for different scopes. Following a similar logic, the term Quantum Processing Unit (QPU) was recently established to identify the computational capability offered by QC devices.

The advantages of integrating HPC and QC into a hybrid HPC+QC infrastructure come in many forms. The following points delineate some of the prevalent use--cases of an HPC+QC system:

\begin{enumerate}[label=(\alph*)]
    \item\textbf{Hybrid classical--quantum algorithms}: Quantum circuits often feature a modular structure, with parameterized gates that are modifiable through classical inputs. Variational Quantum Algorithms (VQA) -- such as the Variational Quantum Eigensolver (VQE) for chemistry problems~\cite{Peruzzo2014}, and the Quantum Approximate Optimization Algorithms (QAOA) for combinatorial optimization problems~\cite{farhi2014quantum} -- require the interplay with classical routines to be tuned to the specific objective. In this scenario, classical computers can run parallel parameter optimization using one or more optimizers, whilst quantum computers are used to prepare quantum states and measure observables, which are then fed back into the classical optimization loop running on a classical resource. Furthermore, as the relationship between the classical parameters and the cost function is non--linear, the optimization landscape of VQA problems can be highly complex with many local minima, and, even in the case of a single optimizer, providing convergence to acceptable values can be challenging for ordinary computers.
    
    \item\textbf{QC supports HPC}: In the realm of machine learning, several hybrid algorithms have been developed that are mainly classical but could benefit from having a quantum component. A Quantum Support Vector Machine (QSVM)~\cite{rebentrost} is a classifier that utilizes a quantum kernel to divide the dataset. In such case, having an HPC infrastructure available allows high--performance pre--processing of large datasets and faster execution time, and to promptly compare the results obtained using a classical kernel. In Quantum Generative Adversarial Networks (QGANs)~\cite{qgan}, whilst HPC trains the discriminator and optimize the overall network, the quantum computer runs the quantum generator, exploring the high--dimensional space in order to generate new data samples that can be used in the discriminator. In a similar way, for deep learning, in Quantum Convolutional Neural Networks (QCNNs)~\cite{Cong2019}, whilst HPC performs the heavy computational tasks associated with training the classical layers, the quantum computer takes care of implementing the quantum layers. 

    \item\textbf{HPC supports QC}: In addition to the classical processing required for variational hybrid algorithms, HPC resources are crucial for extracting the most out of quantum computers through pre-- and post--processing of data. Pre--processing includes optimal compiling and transpiling of circuits, and increasingly, creating the quantum algorithms in the first place.\cite{nvidia-quantum-llm} Efficient compiling and transpiling is necessary in order to make the final executable circuit sufficiently shallow for NISQ devices to finish before decoherence. With increasing qubit count, qubit routing and gate optimization becomes a computationally hard problem, requiring supercomputing resources. Post--processing includes error mitigation measures, in effect, a process of enhancing the signal--to--noise ratio. Also this task becomes increasingly resource--intensive as qubit number grows. In order to avoid losing the computational advantage of a conceptually efficient quantum algorithm, it is important that the pre-- and post--processing tasks can be performed efficiently, that is, with low scaling with respect to qubit count. Here, machine learning techniques are expected to play a decisive role in suppressing the scaling of classical processing methods.\cite{Algorithmiq-TEM}
    
    \item\textbf{HPC simulates QC}: HPC is crucial for benchmarking quantum computers, in assessing the current capabilities and scalability of quantum algorithms, and predicting their future potential. On a classical computer, the exact simulation of a quantum circuit of $n$ qubits requires storing the complex amplitudes for $2^n$ possible states. As the memory necessary to accommodate these values grows exponentially with circuit size, the evaluation of circuits with around $25$ and more, quickly becomes unfeasible for an ordinary machine. Instead, a supercomputer can take advantage of a distributed memory approach, distributing the complex values to different nodes, thereby pushing the boundaries of classical simulation around the $50$ qubits mark, thus enabling larger scalability analysis. 
    Conversely, approximate evaluations of locally--entangled quantum circuits can be classically performed using tensor network methods, which encode and compress a circuit, alleviating the aforementioned memory constraint. Depending on the specific circuit and the tensor structure chosen, it is sometimes possible to approximate circuits with a large number of qubits, even hundreds or thousands, provided that the entanglement is low enough~\cite{patra2023efficient}. However, in such cases, in order to reach the final result, several tensor contractions need to be performed and an efficient contraction path needs to be found, shifting the computational burden from memory to processors~\cite{Gray_2021}. In conclusion, HPC is still indispensable in providing the necessary computational power for these simulations \cite{kicked}.  
\end{enumerate}

\begin{figure}[H]
    \centering
    \includegraphics[width=\textwidth]{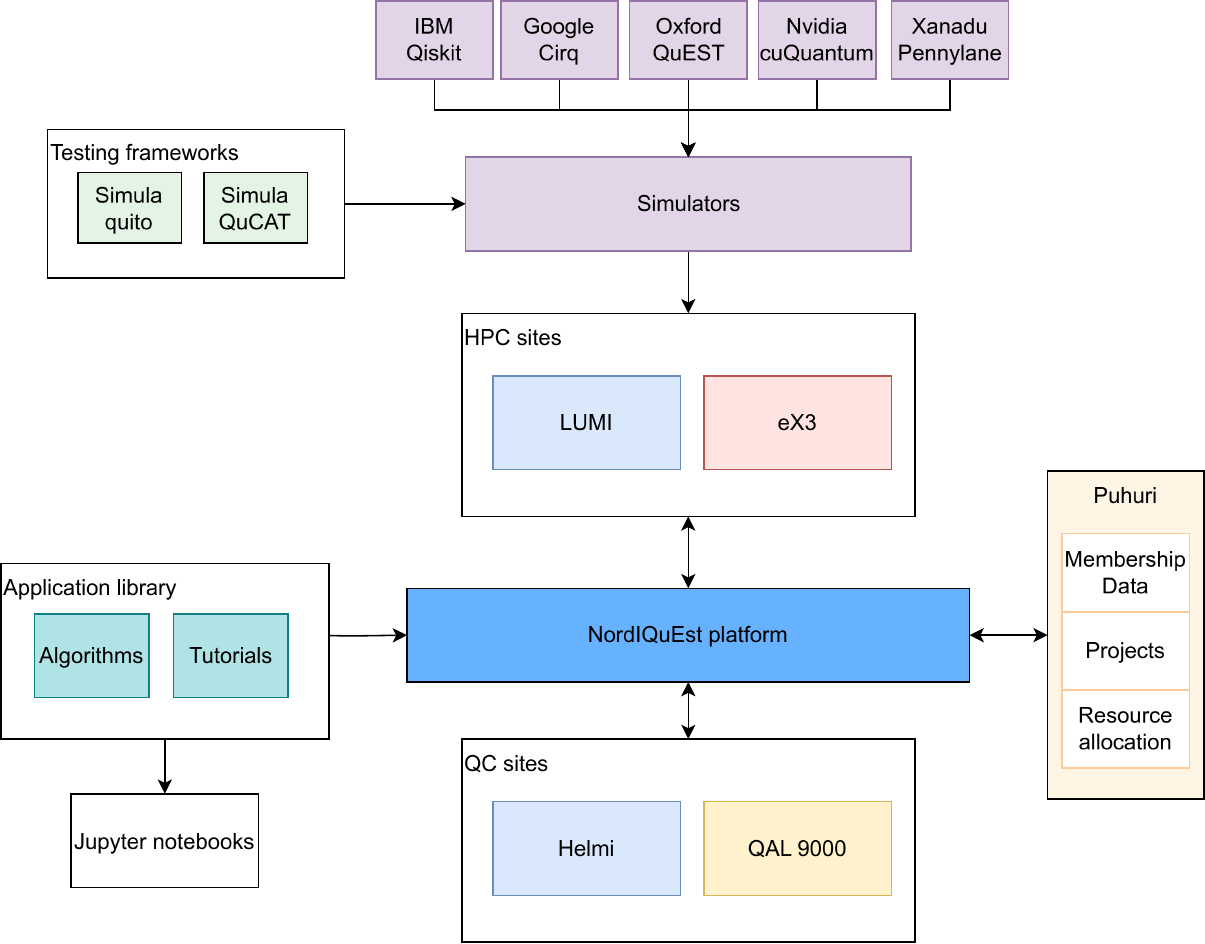}
    \caption{A diagram of the NordIQuEst infrastructure, showing how the different components are connected to each another. The colors of the HPC and QC sites boxes reflect the location of the hardware: Finland (light blue), Norway (red), Sweden (yellow).}
    \label{fig:nq}
\end{figure}
    \section{NordIQuEst infrastructure}
\subsection{Overview}
The NordIQuEst platform acts as the central actor to manage the interaction between HPC systems and QC devices, as shown in Fig.\ref{fig:nq}. The infrastructure is realized through the integration of several hardware and software components, which NordIQuEst enables to seamlessly interface:

\begin{enumerate}[label=(\alph*)]
\item \textbf{HPC sites}: The pan--European EuroHPC LUMI\footnote{\href{https://www.lumi-supercomputer.eu/}{https://www.lumi-supercomputer.eu/}}, is presently the fastest supercomputer in Europe and number five on the global TOP500 list\footnote{\href{https://www.top500.org/}{https://www.top500.org/}}. LUMI consists of 2048 LUMI--C CPU nodes, each with two AMD EPYC 7763 CPUs with 64 cores, 2978 LUMI--G nodes, each with four AMD MI250x GPUs, and additional specialized computing partititons. All LUMI compute nodes use the HPE Cray Slingshot--11 200 Gbps network interconnect. LUMI has a sustained computing power of 380 PFLOPS, and is equipped with $1.75$ PB of RAM. It is hosted by CSC -- IT Centre for Science (Finland), and is the principal HPC resource of NordIQuEst. In addition, the eX3\footnote{\href{https://www.ex3.simula.no/}{https://www.ex3.simula.no/}} project, coordinated by Simula (Norway), provides several different computational clusters for Norwegian researchers. HPC sites are equipped with the SLURM workload manager\footnote{\href{https://slurm.schedmd.com/}{https://slurm.schedmd.com/}}, to schedule jobs and dispatch them to QC sites.

\item \textbf{QC sites}: Helmi\footnote{\href{https://vttresearch.github.io/quantum-computer-documentation/helmi/}{https://vttresearch.github.io/quantum-computer-documentation/helmi/}}, a 5--qubit superconducting quantum computer, hosted at VTT Research Center (Finland), is the main QC site for NordIQuEst. The qubits are arranged in a star--topology and have fidelities around 99.7\% for single--qubit gates, 95\% for two--qubit gates, and 95\% for readout, with T2 times in the range of 10--20$\mu$s. Of a similar superconducting technology, QAL 9000\footnote{\href{https://www.qal9000.se/}{https://www.qal9000.se/}}, is currently a 25--qubit testing chip, with planar grid connectivity map, fabricated and mantained by Chalmers--WAQCT (Sweden). Both QC sites are planning to upgrade to a 40--50 qubits device in the upcoming years.

\item \textbf{User management}: Puhuri\footnote{\href{https://puhuri.io/}{https://puhuri.io/}} is a cloud service project, funded by NeIC, which provides identity management, project management and resource allocation across Europe. It is integrated into both HPC and QC systems.

\item \textbf{Programming frameworks}: Several QC simulators are available, in order to accommodate different needs. IBM Qiskit\footnote{\href{https://www.ibm.com/quantum/qiskit}{https://www.ibm.com/quantum/qiskit}} and Google Cirq\footnote{\href{https://quantumai.google/cirq}{https://quantumai.google/cirq}} are popular well--supported and well--equipped quantum programming frameworks. Oxford QuEST\cite{Jones2019} is specially made for running on multi--core and multi--node clusters, in addition to supporting distributed memory. Nvidia cuQuantum\cite{bayraktar2023cuquantum} is designed for running on GPUs and performing advanced tensor networks calculations. Xanadu Pennylane~\cite{bergholm2022pennylane} contains several tools to run quantum machine learning algorithms and is suited for running large scale hybrid algorithms on supercomputers~\cite{asadi2024hybrid}.

\item \textbf{Testing frameworks}: In addition to the simulators, customized Python software for circuit testing is made available and maintained by Simula: quito\footnote{\href{https://github.com/Simula-COMPLEX/quito}{https://github.com/Simula-COMPLEX/quito}}, an automatic test coverage generator tool, and QuCAT\footnote{\href{https://github.com/Simula-COMPLEX/qucat-tool}{https://github.com/Simula-COMPLEX/qucat-tool}}, a quantum circuit analyzer tool. 

\item \textbf{Application library}: A quantum algorithm library, written in Python, is provided, in order to test and evaluate the capabilities of the hybrid computing framework. Introductory tutorials and additional educational resources are publicly available for new researchers and are easily executable through Jupyter notebooks.
\end{enumerate}

\subsection{\label{sec:workflow}Operational workflow}
\begin{figure}[H]
    \centering
    \includegraphics[width=\textwidth]{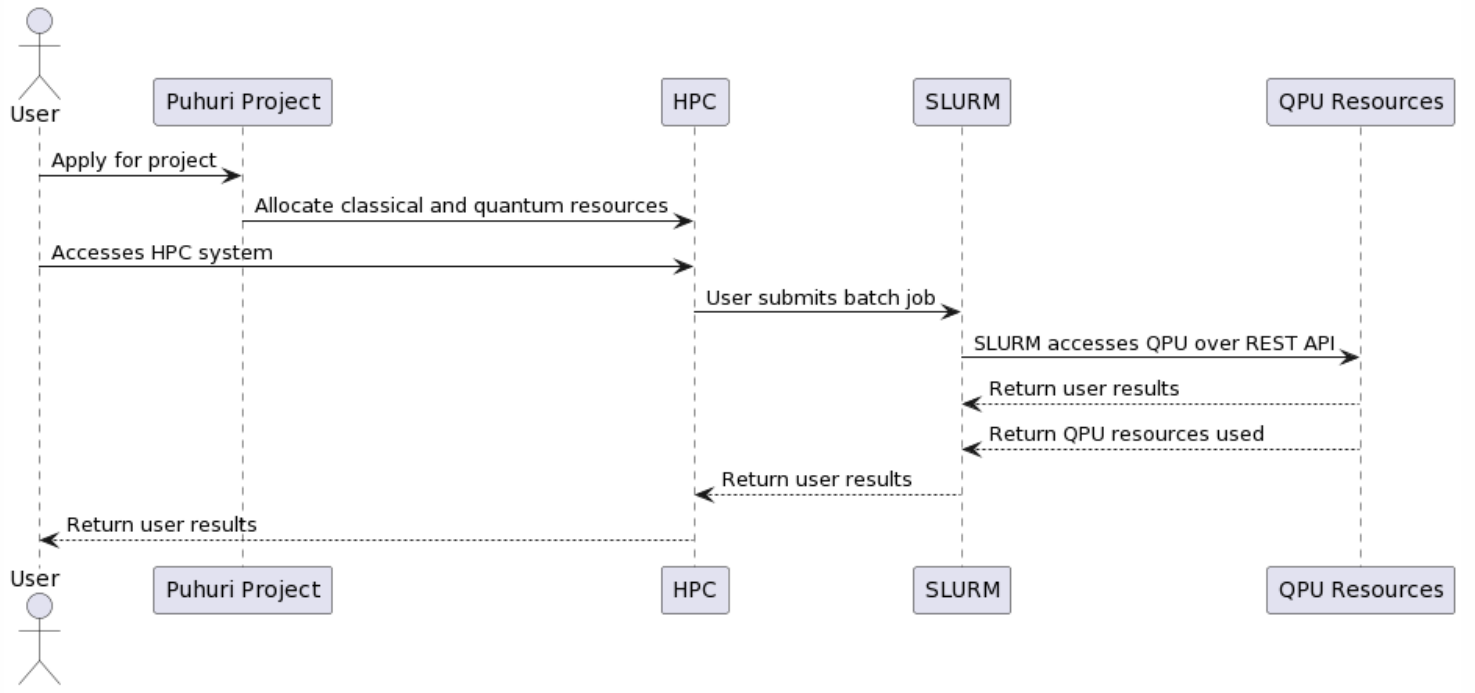}
    \caption{A diagram of a typical user experience on the NordIQuEst platform.}
    \label{fig:workflow}
\end{figure}

The operational workflow designed within the NordIQuEst infrastructure, displayed in Fig. \ref{fig:workflow}, is constructed to be intuitive, and to allow users to easily submit computational jobs to either HPC clusters, QC devices, or a combination of both. Initially, the user applies for a project through Puhuri. Upon approval, Puhuri allocates the necessary classical and quantum resource quotas. The user then accesses the HPC system directly. For job submission, the user sends single or a batch of jobs to the HPC, which are handled by the SLURM workload manager. SLURM manages the job execution, based on the job requirements and resource availability, including accessing the QC devices via a REST API for hybrid computations. The assigned QC device processes the jobs and returns the results and resource usage information to SLURM, which then relays both the results and the metadata to the HPC system. Finally, the HPC system returns the computational results to the user, and reports the resource usage back to Puhuri for project accounting. This workflow ensures an efficient allocation of resources based on the job requirements, availability, and user entitlements, whilst managing the intricate process of coordinating between HPC and QC devices.

\subsection{User engagement}
The NordIQuEst project kicked off officially in April 2022, and the infrastructure became operational in a testing phase in September 2022, in conjunction with the first NordIQuEst workshop.  
Since November 2022, a stable connection between LUMI and the VTT quantum computer Helmi has allowed for testing the HPC+QC infrastructure across the Nordics and Estonia. For example, the connection has been utilised in NordIQuEst organised events, such as the Quantum Autumn School 2023, co--organised with the EuroCC National Competence Centre Sweden (ENCCS)\footnote{\href{https://enccs.se/}{https://enccs.se/}} and the Wallenberg Centre for Quantum Technologies. It is also used at the NeIC Conference 2024, Nordic e--Infrastructure Tomorrow, within the theme of these Proceedings. 

The setup of the NordIQuEst platform has been shown to be stable and able to handle large loads, as tested during the Quantum Autumn School 2023, which is highlighted in Table. \ref{tab:users}. This shows the usage statistics over the short period of the event, where users were introduced to hybrid HPC+QC concepts, and were able to run their first hybrid HPC+QC jobs. The success of this event was aided by the Finnish Quantum--Computing Infrastructure (FiQCI),\footnote{\href{https://fiqci.fi/}{https://fiqci.fi/}} which is one of the main synergistic efforts working together with the NordIQuEst project. The table displays the total number of users which used the platform over the course of the event, with the total number of jobs submitted and the number of shots used. The high number of shots, which is a single execution of an algorithm on the QPU, highlights the engagement of the users during this event.

\begin{table}[H]
\centering
\begin{tabular}{|c|c|c|}
\hline
Users & Total jobs submitted & Shots used \\ \hline
83 & 364 & 2533588 \\ \hline
\end{tabular}
\caption{Statistics of HPC+QC jobs run for the Quantum Autumn School 2023, from October 25th 2023 to October 28th 2023.}
\label{tab:users}
\end{table}

    \section{Challenges: present and future}
The NordIQuEst infrastructure, while pioneering in its integration HPC and QC, has faced a multitude of diverse challenges, as expected for a conceptually completely new implementation. These challenges encompass technical, operational, and conceptual domains, reflecting the complexity of merging these powerful but different computing paradigms. This chapter identifies the most prominent of the challenges faced, and offers some insights into conceiving potential solutions.

\subsection{Technical challenges}
\begin{enumerate}[label=(\alph*)]
\item \textbf{Quantum hardware variability}: The current landscape of quantum computing is full of varying emerging technologies, such as superconducting, ion trap and neutral atoms, each with its own unique strengths and limitations. This diversity has lead to significant variability among available quantum computers in terms of qubit count, error rates and operational capabilities. 

Addressing this challenge requires the development of an adaptable infrastructure, to match computational tasks to the most suitable quantum hardware. The NordIQuEst project has recognized the uncertainty surrounding quantum technology and aims to mitigate its impact by abstracting the hardware layer as much as possible. QC sites in the NordIQuEst project try and offer as much information about the quantum computing stack as possible to the HPC sites through a variety of cutting edge software solutions. Additionally, by offering a wide variety of simulators, with the classical resources to back it up, users can experiment and test out potentially different solutions. 

\item \textbf{Quantum hardware reliability}: Current NISQ quantum computers are delicate devices, requiring lots of calibration work and human maintenance to maintain uptime. This is in stark contrast to the HPC machines with which they are connected to, which provide users with very high up--times, availability and performance. Within the NordIQuEst project, it is the responsibility of QC sites to take care of maintenance and upkeep of the quantum computers, in addition to providing HPC sites with the availability information, such as whether the quantum computer is available for job submission and what is the current calibration status. Presently, for LUMI, this is solved in a straightforward manner with the automatic opening and closing of the SLURM job scheduler upon signal from Helmi. 

\item \textbf{Software instability}: The software enabling the use of quantum computation, be it simulator or real device, is in a very unstable state, reflecting the current state of quantum computing. Within the confines of a cross--border project it has proven to be a difficult challenge to maintain and update software packages, so that different software stacks are in sync and interoperable. The NordIQuEst project has learned that software installers are required to keep the software updated and to test new versions frequently. In addition, researchers are in close contact with maintainers in order to track potential problems and fix them promptly. 
\end{enumerate}

\subsection{Operational challenges}
\begin{enumerate}[label=(\alph*)]
    \item \textbf{Quantum hardware availability}: 
    Quantum computers are typically designed to be used by one user at a time, utilizing a first--in--first--out (FIFO) queuing system at the level of the control electronics and lab equipment. Due to this, access to the quantum device is sequential, and notably, users may be accessing the device from different sources. Additionally, there needs to be a mechanism for users to reserve time slots for their quantum algorithms, to ensure exclusive access and the best use of their time. A sufficiently smart enough scheduler could be employed to manage the execution of quantum jobs which avoids stalling the quantum computer and efficiently plans future job execution. The scheduler should consider factors such as the current quality of the device, time constraints, and the demand for the quantum resources. An early solution, adopted within the FiQCI framework, was to set a daily time slot for HPC users and implement a hard limit to the number of job executions and time limit of jobs to avoid stalling. This required good communication between HPC and QC sites. The future hope is to have a dedicated quantum metascheduler, with a fair--share queuing system, that takes into account both the HPC system and the quantum computer. The implementation of an efficient co--scheduler is expected to be one of the most challenging tasks for HPC+QC infrastructures.
    
    \item \textbf{User quotas and project management}: The administrative tasks that are required for HPC style projects can be seen as a hindrance for access to experimental resources such as quantum computers. Estimating the amount of classical and quantum resources needed is difficult for researchers of hybrid quantum algorithms and complex processes for project management might prevent adoption from new researchers. User quotas impose restrictions on the amount of computational resources that individual researchers or projects can access. Experienced researchers with demanding computational requirements may find themselves constrained by these quotas, leading to delays in their work. To address these challenges organizations can organize workshops to deliver temporary collective access. While the practical experience among the user base is building up, it is important to provide a flexible resource application process, with short application processing times. Additionally, offering training through easy to follow tutorials and Jupyter notebooks can help new researchers familiarize themselves with quantum computing concepts and the tools needed to perform hybrid computation. 
\end{enumerate}

\subsection{Conceptual challenges}
\begin{enumerate}[label=(\alph*)]
    \item \textbf{Hybrid applications assessment}: Properly evaluating quantum algorithms is a challenge due to the many heuristics involved. The nature of hybrid algorithms combining both classical and quantum components leads to additional complexity in terms of assessing how well they are performing, how well they scale, and the potential applications. The NordIQuEst project has found that many users prioritize the execution of quantum algorithms over in--depth analysis of benchmarking, testing the scalability and assessing the application. For many, it is important that it just executes. These challenges can be addressed by promoting evaluation studies that focus on the assessment of hybrid quantum applications. Providing software tools that facilitate the isolation of variables and automate the assessment of algorithms. An important component is to enable mitigation of the noise from NISQ--era quantum devices to enhance the reliability and reproducibility of results. Classical resources in HPC environments are in an abundance compared to their quantum counterparts. Hybrid applications should try to leverage the parallel computing power offered by HPC to enhance the assessment of hybrid algorithms. 
    
    \item \textbf{User engagement} While the NordIQuEst infrastructure is still under development and not generally available to users, it is important to proactively consider user engagement. Our experience is that some clear barriers to user adoption exist. Some of these are related to insufficient knowledge about what an HPC+QC infrastructure can provide for a user, and others to the practicalities of getting access to it. As discussed in Sec.~\ref{sec:workflow}, the envisioned journey of a prospective user begins with applying for a project. This requires that the user articulates their research objectives in order to properly assess the computational quotas. Although necessary, this process might be hold back new researchers in adopting experimental technologies. As we have ascertained during NordIQuEst's events, facilitating this by providing access to the platform through a single project account and sharing its resources, significantly lowers this barrier. Especially for users who are not familiar with standard HPC procedures, providing a hands--on walk--through of the process is highly beneficial. Conversely, more experienced QC users might already have direct access to QC devices -- such as those hosted by VTT or Chalmers -- and might not resort to NordIQuEst's services unless they see an added value, e.g., in the form of hybrid computation or heavy pre-- and post--processing. This added value thus needs to be clearly formulated and communicated to potential users. To engage users, it is also crucial to build up the infrastructure in constant rapport with the end--users. In practice, this means that the project implementation needs to be sufficiently flexible to allow for developing the user experience on the fly. 
    
    \item \textbf{Education and training}: Both HPC and QC are complex domains that demand a level of expertise and experience to navigate efficiently. Integrating HPC with QC adds an additional layer of complexity, resulting in a steep learning curve for new users. Becoming proficient in both HPC and QC requires dedicated time and education in quantum computing, using a supercomputer, computer science, and additional domains. Bridging the gap between the HPC realm and the QC realm is crucial for enabling the adoption of hybrid infrastructure. This requires developing training programs and educational material tailored to equip researchers and potential users with the necessary skills and knowledge. Such training programs should include primers, tutorials, and plug--and--play demonstrations for both beginners and advanced users. In particular, the NordIQuEst project has seen the need for plug--and--play examples to introduce users and increase adoption. Advanced examples of hybrid algorithms are also needed for experienced users looking to optimize their workflows. Such examples should demonstrate the utilization of open--source tools and cutting--edge techniques to get the most out of the resources available. 
\end{enumerate}
    \section{Conclusions}
The NordIQuEst collaboration has established a pioneering groundwork for integrating HPC and QC resources into a cohesive, hybrid computing infrastructure. This effort -- the first of its kind on such a scale -- was driven by the potential of offering researchers the opportunity to experiment with the novel technology of quantum computing, while accessing the capabilities of usual supercomputers. 

Since its inception, the NordIQuEst platform has been predisposed with a diverse set of tools, enabling researchers to advance important scientific goals, such as evaluating variational hybrid algorithms, benchmarking current quantum computers, and implementing new hybrid applications. Nonetheless, the NordIQuEst project has faced, and successfully addressed, several technical challenges, including the need for an adaptable infrastructure that can accommodate the diverse nature of quantum hardware, and the software instability inherent in this rapidly evolving field. On the operational side, the project has managed to provide reliable access to quantum hardware, while leveraging existing systems and collaborative projects such as FiQCI (Research Council of Finland) and Puhuri (NeIC/Nordforsk) for coordinating user access and resource allocation. Furthermore, the project's commitment to cultivating a research community, has produced hands--on training events and educational materials, aimed at bridging the knowledge gap between classical HPC users and the QC practitioners. 

In conclusion, while the NordIQuEst project has made significant strides in unifying HPC and QC systems, it is only an initial step towards a future where the synergistic potential of these technologies can be fully realized. Our journey underscores the importance of international collaboration in fostering innovation at the forefront of computing technologies, setting a precedent for future endeavors that will build on our foundational work. We end by highlighting the need for open--source development and open access material for enabling adoption and (re)utilization of the work performed.
    \section{Acknowledgments}

We thank the NeIC/Nordforsk collaboration for providing funding and believing in the NordIQuEst project. We further appreciate the operational support offered by the Puhuri project. We acknowledge the invaluable contribution of every institution participating in the NordIQuEst collaboration. 
    
    %
    %
    \bibliographystyle{cls/splncs04}
    \bibliography{refs}

\end{document}